\documentclass[conference]{IEEEtran}

\usepackage{cite}
\usepackage{amsmath,amssymb,amsfonts}
\usepackage{algorithmic}
\usepackage{graphicx}
\usepackage{textcomp}
\usepackage{xcolor,colortbl}
\usepackage{rquestion}
\usepackage{mdframed}

\usepackage{hyperref}
\hypersetup{
    colorlinks,
    linkcolor={black},
    citecolor={black},
    urlcolor={black}
}
\usepackage{cleveref}
\usepackage{balance}

\newcommand{\explain}[2]{%
\noindent\emph{\underline{#1}} #2}
\newcommand{\why}[1]{\explain{Why:}{#1}}
\newcommand{\how}[1]{\explain{How:}{#1}}
\newcommand{\finding}[1]{}

\usepackage{listings}

\definecolor{mygreen}{rgb}{0,0.6,0}
\definecolor{lightgreen}{rgb}{0.6,0.9,0.6}
\definecolor{lightyellow}{rgb}{0.9,0.9,0.6}
\definecolor{lightorange}{rgb}{0.9,0.8,0.6}
\definecolor{lightred}{rgb}{0.9,0.7,0.7}
\definecolor{mygray}{rgb}{0.5,0.5,0.5}
\definecolor{lightgray}{rgb}{0.8,0.8,0.8}
\definecolor{mymauve}{rgb}{0.58,0,0.82}

\usepackage{pifont}

\newcounter{lstannotation}
\renewcommand{\thelstannotation}{\ding{\number\numexpr181+\arabic{lstannotation}}}
\newcommand{\annotation}[1]{\refstepcounter{lstannotation}\label{#1}\thelstannotation}

\newcommand{\myurl}[1]{\url{#1} (Visited on \today)}

\usepackage{array,tabularx}

\newenvironment{conditions}
  {\par\vspace{\abovedisplayskip}\noindent
   \tabularx{\columnwidth}{>{$}l<{$} @{${}={}$} >{\raggedright\arraybackslash}X}}
  {\endtabularx\par\vspace{1.3\belowdisplayskip}}

\newcounter{BCHCounter}
\newcommand{\bchInstance}[6]{%
\refstepcounter{BCHCounter}%
\begin{mdframed}[leftline=false,rightline=false]
{\centerline{\textsc{Maintainability Instance \theBCHCounter}}}
\end{mdframed}
\noindent \emph{\textbf{Git Repository:}} \url{https://github.com/#1}\\
\emph{\textbf{Commit:}} \href{https://github.com/#1/commit/#2}{\texttt{#2}}\\
\emph{\textbf{Change:}} #5\\
\emph{\textbf{Maintainability issue ($\Delta M=#4$):}} #6\par\vspace{-2ex}\noindent\rule{\linewidth}{0.2pt}}

\usepackage{todonotes}

\graphicspath{{./figures},{.}}

\renewcommand{\baselinestretch}{0.97}
\renewcommand{\baselinestretch}{0.982}
\begin{document}
\title{Do Energy-oriented Changes Hinder Maintainability?}

\author{
\IEEEauthorblockN{Luis Cruz\IEEEauthorrefmark{1},
Rui Abreu\IEEEauthorrefmark{2},
John Grundy\IEEEauthorrefmark{3},
Li Li\IEEEauthorrefmark{3} and
Xin Xia\IEEEauthorrefmark{3}}
\IEEEauthorblockA{\IEEEauthorrefmark{1} INESC-TEC, University of Porto, Porto, Portugal\\
Email: luiscruz@fe.up.pt}
\IEEEauthorblockA{\IEEEauthorrefmark{2} INESC-ID, University of Lisbon, Lisbon, Portugal\\
Email: rui@computer.org}
\IEEEauthorblockA{\IEEEauthorrefmark{3} Faculty of Information Technology, Monash University, Melbourne, Australia\\
Email: \{john.grundy, li.li, xin.xia\}@monash.edu}\vspace{-3em}
}

%
%

%

\maketitle

\begin{abstract}

Energy efficiency is a crucial quality requirement for mobile applications.
However, improving energy efficiency is far from trivial as developers lack the
knowledge and tools to aid in this activity. In this paper we study the impact
of changes to improve energy efficiency on the maintainability of Android
applications. Using a dataset containing 539 energy efficiency-oriented
commits, we measure maintainability -- as computed by the Software Improvement
Group's web-based source code analysis service \emph{Better Code Hub} (BCH) --
before and after energy efficiency-related code changes. Results show that in
general improving energy efficiency comes with a significant decrease in
maintainability. This is particularly evident in code changes to accommodate
the \emph{Power Save Mode} and \emph{Wakelock Addition} energy patterns. In
addition, we perform manual analysis to assess how real examples of
energy-oriented changes affect maintainability. Our results help mobile app
developers to 1) avoid common maintainability issues when improving the energy
efficiency of their apps; and 2) adopt development processes to build
maintainable and energy-efficient code. We also support researchers by
identifying challenges in mobile app development that still need to be
addressed.

\end{abstract}

\begin{IEEEkeywords}
Energy Consumption, Software Maintenance, Mobile Computing
\end{IEEEkeywords}
\vspace{-1em}


\section{Introduction}
\label{sec:intro}

Modern mobile applications, popularly known as apps, provide users with a
number of features in multi-purpose mobile computing devices -- smartphones.
The convenience of using smartphones to pervasively accomplish important daily
tasks has a big limitation: smartphones have a limited battery life. Apps that
drain battery life of smartphones can ruin user experience, and
are likely to be uninstalled unless they offer a key feature.

Thus, it is critically important that apps efficiently use the battery of
smartphones. However, many developers still lack knowledge about best practices
to deliver energy efficient mobile
applications~\cite{pang2015programmers,sahin2014code}. Important efforts have
been carried out to help developers ship energy efficient mobile
apps~\cite{kong2018automated}. Novel tools have been built to suggest energy
improvements to the codebases of mobile
apps~\cite{cruz2018using,cruz2018measuring,linaresvasquez2018multiobjective,li2016automated} and to help developers measure the energy consumption of
their
apps~\cite{chowdhury2018greenscaler,boonkrong2015reducing,chowdhury2016greenoracle,di2017petra,li2017static}.

Despite these efforts, improving the energy efficiency of mobile applications
is not a trivial task. It requires implementing new features and refactoring
existing ones~\cite{cruz2019catalog}, only for the sake of better energy usage,
i.e., predominantly a non-functional rather than functional change. However,
the extent to which these changes affect the maintainability of the mobile app
software has not yet been studied. In this work, we are interested in studying
the trade-off between the energy efficiency and the maintainability of mobile
applications.

The International Standards on software quality ISO/IEC~25010 define software
maintainability as ``the degree of effectiveness and efficiency with which a
software product or system can be modified to improve it, correct it or adapt
it to changes in environment, and in requirements''~\cite{iso25010}. The
standard defines five core sub-characteristics of maintainability: modularity,
reusability, analyzability, modifiability, and testability. The Software
Improvement Group (SIG) has developed a web-based source code analysis toolset
\emph{Better Code Hub} (BCH)~\cite{visser2016building} that maps the
ISO/IEC~25010 standard on maintainability into a set of 10 guidelines, such as
\emph{write short units of code} and \emph{write code once}, derived from
static analysis~\cite{kuipers2007practical,baggen2012standardized,visser2016building,olivari2018maintainable}.
The code metrics used by the SIG model were
empirically validated in previous work~\cite{bijlsma2012faster}. We use this
toolset in our work to provide an assessment of maintainability in mobile
app codebases.

Specifically, we want to explore whether there is a trade-off between applying energy
efficiency patterns and keeping the maintainability of the apps, i.e., does
improving energy efficiency have a negative impact on code maintainability? In
this paper, we present the results of our analysis on the maintainability using
539 energy commits harvested from open source Android applications.

The key contributions of this work are:
\vspace{-0.3em}
\begin{itemize}
  \item An empirical investigation of the impact of energy patterns in code maintainability.
  \item A dataset of energy commits and respective impact on maintainability.
  \item A software package with all scripts used in our experiments and a dataset of energy commits with respective impact on maintainability, for reproducibility. Available here: \url{https://figshare.com/s/989e5102ae6a8423654d}.
\end{itemize}
\vspace{-0.3em}
Our empirical study finds evidence that energy efficiency-oriented code changes
have a negative impact on code maintainability. In particular, careful thinking
is required to implement the energy patterns \emph{Power Save Mode} and
\emph{Wakelock Addition}. Furthermore, we show that energy patterns are more
likely to require maintenance than regular code changes.

This paper is structured as follows. In Section~\ref{sec:example}, we introduce
an example of an energy improvement from a real-world mobile application.
Section~\ref{sec:methodology} describes the methodology we use to answer the
research questions. We present the results in Section~\ref{sec:results} and
discuss their implications in Section~\ref{sec:discussion}. In
section~\ref{sec:t2v}, we enumerate the threats to the validity of our work.
Section~\ref{sec:rw} describes the differences between our work and existing
literature. Finally, in Section~\ref{sec:conclusions} we summarize the main
conclusions and elaborate on future work.

\vspace{-0.5em}
\section{Motivating Example \& Research Questions}
\label{sec:example}

%



Improving energy efficiency of apps revolves around changing their codebases.
Previous work has studied existing energy patterns for mobile
applications~\cite{cruz2019catalog}. It cataloged typical coding practices
developers adopt to address energy efficiency. An example of an energy pattern
is the \emph{Power Save Mode}: the app features a mode that can be activated
upon low battery and uses fewer resources while providing the minimum
functionality that is indispensable to the user.

An instance of this pattern can be found in the app
\emph{NetGuard}\footnote{More information about the app NetGuard on Google Play
app store:
\myurl{https://play.google.com/store/apps/details?id=eu.faircode.netguard&hl=en}
 .} -- an Android app that provides a firewall and monitors network traffic
across other apps.

To improve energy efficiency, \emph{NetGuard}'s developers decided to implement
the pattern \emph{Power Save Mode}~\cite{cruz2019catalog}.
The following snippet presents the required code changes\footnote{Commit taken from NetGuard project's Github repository,
available
at: \myurl{https://github.com/M66B/NetGuard/commit/2e70a038970d6efe9f74e5719e7648f91de30498}}:

\setcounter{lstannotation}{0}
\begin{lstlisting}
public class SinkholeService extends VpnService {
  private boolean powersaving = false;
// [snip]

  public void handleMessage(Message msg) {
+  if (powersaving) return;/*!\annotation{lst:suppress}!*/
    switch (msg.what) {
      case MSG_PACKET:
      log((Packet) msg.obj, msg.arg1, msg.arg2 > 0);
// [snip]
    }
  }

// [snip]

  private BroadcastReceiver interactiveStateReceiver =
    new BroadcastReceiver() {
    @Override
    public void onReceive(Context context, Intent intent) {
// [snip]
      statsHandler.sendEmptyMessage(
-      Util.isInteractive(this) ? STATS_START : STATS_STOP
+      Util.isInteractive(this) && !powersaving ?
+        STATS_START : STATS_STOP  /*!\annotation{lst:graph}!*/
      );
    }
  };

// [snip]

+ private BroadcastReceiver powerSaveReceiver =
+   new BroadcastReceiver() { /*!\annotation{lst:handler}!*/
+   @Override
+   @TargetApi(Build.VERSION_CODES.LOLLIPOP) /*!\annotation{lst:version1}!*/
+   public void onReceive(Context context, Intent intent) {
+     Log.i(TAG, "Received " + intent);
+     Util.logExtras(intent);
+     PowerManager pm = getSystemService(
+       Context.POWER_SERVICE);
+     powersaving = pm.isPowerSaveMode();
+     Log.i(TAG, "Power saving=" + powersaving);
+     statsHandler.sendEmptyMessage(
+       Util.isInteractive(this) && !powersaving ?
+         STATS_START : STATS_STOP  /*!\annotation{lst:duplicate}!*/
+     );
+   };

// [snip]

  @Override
  public void onCreate() {
    // [snip]
+   if (VERSION.SDK_INT >= VERSION_CODES.LOLLIPOP) { /*!\annotation{lst:subscribe}\annotation{lst:version2}!*/
+     PowerManager pm = getSystemService(POWER_SERVICE);
+     powersaving = pm.isPowerSaveMode();
+     IntentFilter ifPower = new IntentFilter();
+     ifPower.addAction(ACTION_POWER_SAVE_MODE_CHANGED);
+     registerReceiver(powerSaveReceiver, ifPower);
+   }
// [snip]
  }

  // [snip]

  @Override
  public void onDestroy() {
    // [snip]
+   if (VERSION.SDK_INT >= VERSION_CODES.LOLLIPOP) /*!\annotation{lst:version3}!*/
+     unregisterReceiver(powerSaveReceiver); /*!\annotation{lst:unsubscribe}!*/
// [snip]
  }
}
\end{lstlisting}

\vspace{-1em}
\ref{lst:suppress} Disable data logging methods to suppress output.

\ref{lst:graph} Deactivate network speed statistics when \emph{Power Save Mode} is activated.

\ref{lst:handler} An instance of \texttt{BroadcastReceiver} is created to
implement the handler of Power Save Mode events.

\ref{lst:version1} A decorator is used to make sure \emph{Power Save Mode}
changes are only applied to a compatible Android version.

\ref{lst:duplicate} Network speed statistics have to be deactivated upon
different events. This is a duplicate of \ref{lst:graph}.

\ref{lst:subscribe} Subscribe event of \emph{Power Save Mode} activation.

\ref{lst:version2}\ref{lst:version3} A conditional statement is used to ascertain \emph{Power
Save Mode} is only applied to a compatible Android version.

\ref{lst:unsubscribe} Subscribe \emph{Power Save Mode} event.

Although the concept of creating a \emph{Power Save Mode} is relatively simple, this
example illustrates that a number of code changes have to be made that have an adverse impact on code
maintainability. For instance, it requires adding duplicated code and adding
conditional statements to check the version of Android, increasing cyclomatic
complexity. This form of coding goes against some of the guidelines for
building maintainable software~\cite{visser2016building}.

We are concerned that, while improving energy efficiency, developers are
decreasing the maintainability of their projects, and consequently increasing
technical debt. In this work, we use a dataset of energy efficiency-oriented
changes to measure the difference in maintainability incurred in Android
applications when those changes were applied. Therefore, in this work, we want
to answer the following research questions.

\newrquestion{rq:maintainability}{What is the impact of making code changes to
improve energy efficiency on the maintainability of mobile apps?}

\why{Energy efficiency often requires to change codebases and even the features of
a mobile application. If maintainability is not addressed, these improvements
may significantly increase technical debt and require rework during the lifetime of
the project.}

\how{We analyze a combination of previous datasets with 539 energy-oriented
commits. We compute the maintainability score of these commits using the online
tool BCH. We apply the same approach to a dataset of regular commits to use as
baseline and compare results.}

\finding{Energy-oriented commits significantly decrease software
maintainability in open source Android apps.}

\newrquestion{rq:patterns}{Which energy efficiency-oriented code
change patterns are more likely to affect the maintainability of mobile apps?}

\why{Some energy patterns might be more complex to implement than others. By
understanding which patterns are more likely to introduce maintainability
issues, we bring awareness to mobile app and mobile SDK developers of code
changes that require more attention.}

\how{We use the classification of developers activities made in previous work~\cite{cruz2019catalog,moura2015mining,bao2016android,cruz2018using}
to group energy-oriented commits and analyze maintainability independently.}

\finding{Featuring a \emph{Power Save Mode} or \emph{Wakelocks} significantly
introduces maintainability issues.}

%
%

\newrquestion{rq:examples}{What are typical maintainability issues introduced by energy-oriented code changes?}

\why{By using examples of typical maintainability issues in real
energy-oriented commits, practitioners and researchers will have a more tangible
concept of how energy efficiency may hinder maintainability.}

\how{First, we select energy-oriented commits that yielded low maintainability. Then, we manually
inspect these commits and discuss the potential issues entailed by energy
efficiency improvements.}

%
%
%

\vspace{-0.5em}
\section{Methodology} \label{sec:methodology}
\vspace{-0.5em}

\begin{figure}[!t]
\centering
\includegraphics[width=0.9\linewidth]{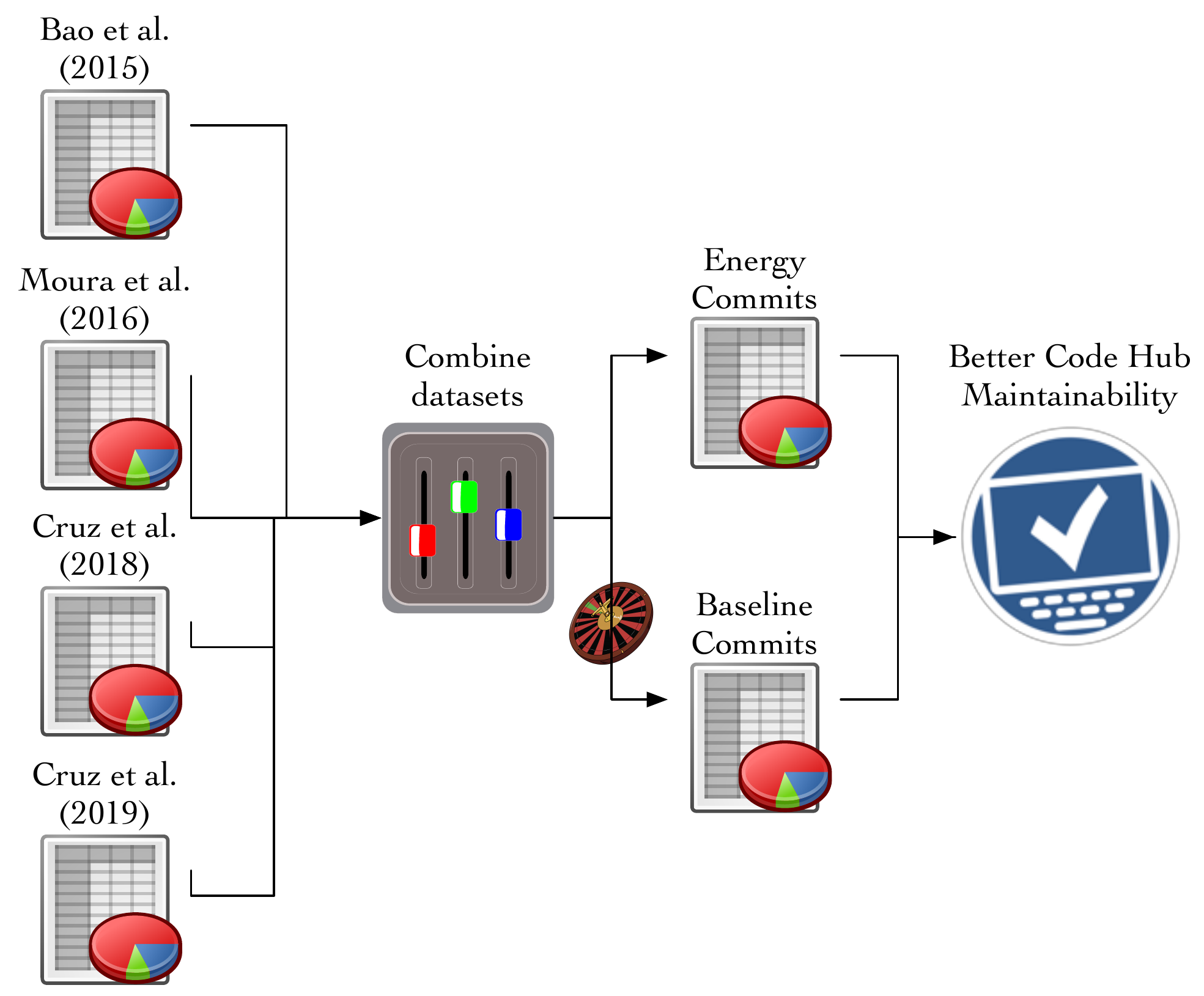}
\caption{Methodology for data collection.}
\label{fig:flow}
\end{figure}

We use the approach illustrated in Fig.~\ref{fig:flow} to analyze how energy
commits affect the maintainability of Android applications. It comprises the
following steps:

\begin{enumerate}

\item Combine the datasets from related work that classifies the activities of
developers addressing energy efficiency in mobile
apps~\cite{bao2016android,moura2015mining,cruz2018using,cruz2019catalog}.

\item Collect regular commits from Android apps to be used as baseline.

\item Compute the impact of energy-oriented commits on maintainability, using BCH.

\end{enumerate}

\subsection{Dataset}

Our work uses the data collected in four previous studies~\cite{cruz2019catalog,moura2015mining,bao2016android,cruz2018using} to
assess the impact of energy management-oriented changes on the maintainability
of Android software.
The datasets are summarized in Table~\ref{tab:datasets} and explained below.

\begin{table}
\caption{Datasets that were combined from previous work.}
\label{tab:datasets}
\begin{tabular}{l c r l}
  \hline
  \rowcolor{lightgray}
Authors & Ref. & \# Commits & Platforms \\
\hline
Moura et al. (2015) &  \cite{moura2015mining} & 2188  & Android, iOS, non-mobile \\
Bao et al. (2016)   &  \cite{bao2016android}  & 468   & Android \\
Cruz et al. (2018)  &  \cite{cruz2018using}   & 59    & Android \\
Cruz et al. (2019)  &  \cite{cruz2019catalog} & 431   & Android, iOS\\
\hline
\end{tabular}
\end{table}

Moura et al. (2015) mined more than $2000$ commits to
understand energy management activities in general-purpose
applications~\cite{moura2015mining}. Their findings suggest that energy
efficiency techniques have to be carefully chosen to ascertain that the
correctness of the software remains intact. In an extension of this
work~\cite{bao2016android}, Bao et al. (2016) used a similar approach to focus
exclusively on Android apps, having mined $468$ energy management commits. They
found that apps in different categories typically have different approaches to
energy efficiency.

Cruz et al. (2018) have provided energy efficiency patterns in an automatic
refactoring tool~\cite{cruz2018using}. The tool was used to analyze 140
open-source Android apps. As an outcome, the authors submitted 59 pull-requests
containing energy improvements to the official repositories of open-source
Android applications. In another work, Cruz et al. (2019) proposed a catalog
with 22 energy patterns to help developers design energy efficient mobile
applications~\cite{cruz2019catalog}. The authors mined the commits, issues, and
pull requests of 1027 Android apps and 726 iOS apps to understand how
developers address energy efficiency issues. The catalog can be used to help
novice developers learn advanced energy management techniques from existing
practices.

From all the data collected, we only select commits from Android projects.
Changes from other platforms, such as iOS and Desktop software, were filtered
out. Moreover, we cleansed the dataset by filtering out projects that have been
deleted and by updating projects that have moved their repositories to a different
location. In addition, datasets~\cite{moura2015mining}~and~\cite{bao2016android}
include commits that have not been manually validated -- we only include
commits that the authors manually ascertained as proper energy changes.

In addition, we reuse the categorization of the energy changes defined in the
original datasets. Despite similar, different datasets use different labels to
indicate the same pattern. For example, the same pattern is labeled as
\textit{PowerConditionalStrategy:PowerSaveMode} by Bao et al.
(2016)~\cite{bao2016android} and as \textit{Power Save Mode} by Cruz et al.
(2019)~\cite{cruz2019catalog}. We map these and other identical categories into
unique labels\footnote{The whole set of identical categories can be found in
the replication package:\myurl{https://figshare.com/s/16397140e8183708d248}.}.
In sum, energy commits are classified into seven categories:

\begin{itemize}

  \item \textbf{Bug Fix \& Code Refinement.} Changes related to fixing energy
  bugs, or refactoring code that already implements energy management features.

  \item \textbf{Power Awareness.} Have a different behavior when the device is
  connected/disconnected to a power station or has different battery levels.

  \item \textbf{Power Save Mode.} Implementation of an energy efficient mode in
  which some features are deactivated to improve better energy usage.

 \item \textbf{Power Usage Monitoring.} Developers add UIs or configurations
  to inform users about the status of the battery and let them make informed
  decisions about their interaction with the application.

  \item \textbf{Wakelock Addition.} Wakelocks are used when apps execute tasks
  that may take longer to execute and need to prevent resources from getting
  into a sleep state (e.g., screen, network, audio, etc.).

  \item \textbf{Wakelock Optimization.} Inappropriate usage of wakelocks may
  incur into unnecessary energy usage. Thus, often developers have to optimize
  wakelock behavior, or even replace them with other techniques (e.g., event
  handlers).

  \item \textbf{Miscellaneous.} This comprises several categories of energy
  commits. Since we perform hypothesis tests to statistically validate results,
  we need to have at least 20 commits per category. Thus, when a category
  comprises less than 20 commits, we label it as \emph{Miscellaneous}.

\end{itemize}

\subsection{Baseline Commits}

Although we want to assess the maintainability of energy commits, there is no
evidence in previous work on how regular commits affect the maintainability of
Android projects. E.g., if energy-oriented commits hinder maintainability, we
need to understand whether this result is in fact different from general
purpose commits. Thus, in parallel with energy commits, we also analyze the
maintainability of all other commits and use these as a baseline to answer
RQ\ref{rq:maintainability} and RQ\ref{rq:patterns}.

The baseline dataset is collected as follows: for each energy commit, we obtain
all the commits of the respective project and randomly select one. In addition,
we randomly select 20 commits to validate that commits are similar in terms of
complexity. By using the dataset of energy commits as input for our baseline
dataset we make sure that differences in maintainability in the two datasets
are not originated by the specificities of different Android projects (e.g.,
different contribution policies, coding guidelines, app categories, etc.).

\subsection{Maintainability Analysis}

We make use of the Software Improvement Group's web-based source code analysis
service \emph{Better Code Hub} (BCH for short\footnote{\emph{Better Code Hub}’s
website available at \myurl{https://www.bettercodehub.com/}}) to collect
maintainability reports of the projects. BCH delivers a maintainability model
based on 10 guidelines~\cite{visser2016building}:

\begin{enumerate}

  \item \textbf{Write short units of code.} Long units are hard to test, reuse,
  and understand.

  \item \textbf{Write simple units of code.} Keeping the number of branch
  points low makes units easier to modify and test.

  \item \textbf{Write code once}. When code is duplicated, bugs need to be
  fixed in multiple places, which is inefficient and prone to errors.

  \item \textbf{Keep unit interfaces small}. Keeping the number of parameters
  low makes units easier to understand and reuse.

  \item \textbf{Separate concerns in modules}. Changes in a loosely coupled
  codebase are much easier to oversee and execute than changes in a tightly
  coupled codebase. This is computed based on the total fan-in of all methods
  in a module. Note that a module in Java and other object-oriented languages
  translates to a class.

  \item \textbf{Couple architecture components loosely}. Independent components
  ease isolated maintenance.

  \item \textbf{Keep architecture components balanced}. Balanced components
  ease locating code and foster isolation, improving maintenance activities.

  \item \textbf{Keep your codebase small}. Small systems are easier to search
  through, analyze, and understand code.

  \item \textbf{Automate tests}. Automated testing makes development
  predictable and less risky.

  \item \textbf{Write clean code}. Code without code smells is less likely to
  bring maintainability issues.

\end{enumerate}

For each guideline, BCH evaluates the compliance against a particular guideline
by setting boundaries for the percentage of code allowed to fall in each of the
four risk severity categories (\textit{low risk}, \textit{medium risk},
\textit{high risk}, and \textit{very high risk}). If the thresholds are not
violated, the project is considered to be compliant with the guideline.
According to BCH, the guideline thresholds are calibrated yearly based on a
representative benchmark of closed and open source software systems. Being
compliant with a guideline means that the project under analysis is at least
better than 65\% of the software systems in BCH's benchmark.

\begin{figure}[!t]
  \centering
  \includegraphics[width=\linewidth]{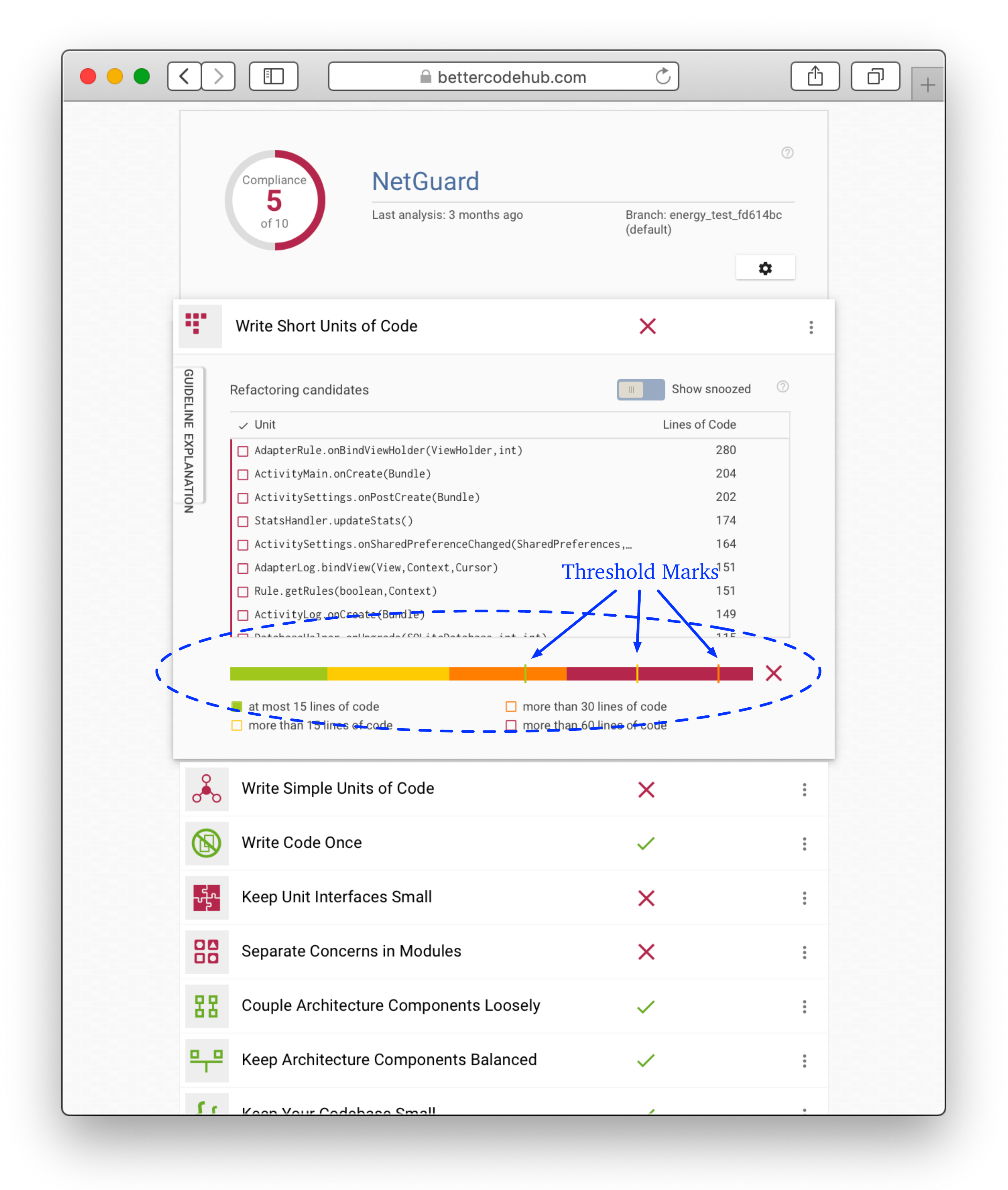}
  \caption{BCH's maintainability report of the app \emph{NetGuard} for the guideline \emph{Write short units of code}. The app does not comply with the guideline because the bars are not reaching the threshold marks.}
  \label{fig:bch_noncompliant}
\end{figure}

The BCH report of the app \emph{NetGuard} for a non-compliant guideline can be
seen in Fig.~\ref{fig:bch_noncompliant}. This was extracted from the report of
the app \emph{NetGuard}, used in the motivating example of
Section~\ref{sec:example}. The green bar represents the percentage of compliant
lines of code. These lines of code are considered to be compliant with ISO
25010 standard for maintainability~\cite{organisation2011systems}. The yellow,
orange and red bars represent non-compliant lines of code with \emph{medium},
\emph{high}, and \emph{very high} severity levels, respectively. Along the
bars, there are also marks that refer to the compliance thresholds for each
severity level. The report is equivalent to the information reported in
Table~\ref{tab:bch_report}: a set of thresholds, number of lines of code (LOC),
and percentage of the project for each severity level. Nonetheless, thresholds
provided by BCH do not sum to 100\%: non-compliant levels are provided in a
cumulative way (e.g., the threshold for the medium level includes high and very
high levels); the compliant-level threshold is the complement of the
medium-level threshold.

\begin{table}
  \centering
  \caption{Example of a BCH report for a non-compliant case.}
  \label{tab:bch_report}
\begin{tabular}{lccc}
\hline
\rowcolor{lightgray}
Level     & Threshold & LOC     & Percentage of Code \\
\rowcolor{lightgray}
          & (\%)      &         & (\%) \\
\hline
\rowcolor{lightgreen}
(Low)     & (56.3)    & (1353)  & (18.6)\\
\rowcolor{lightyellow}
Medium    & 43.7      & 1683    & 23.2\\
\rowcolor{lightorange}
High      & 22.3      & 1622    & 22.4\\
\rowcolor{lightred}
Very High & ~6.9      & 2588    & 35.7\\
\hline
\end{tabular}
\vspace{0.1in}
\end{table}

\begin{figure}[!t]
  \centering
  \includegraphics[width=0.8\linewidth]{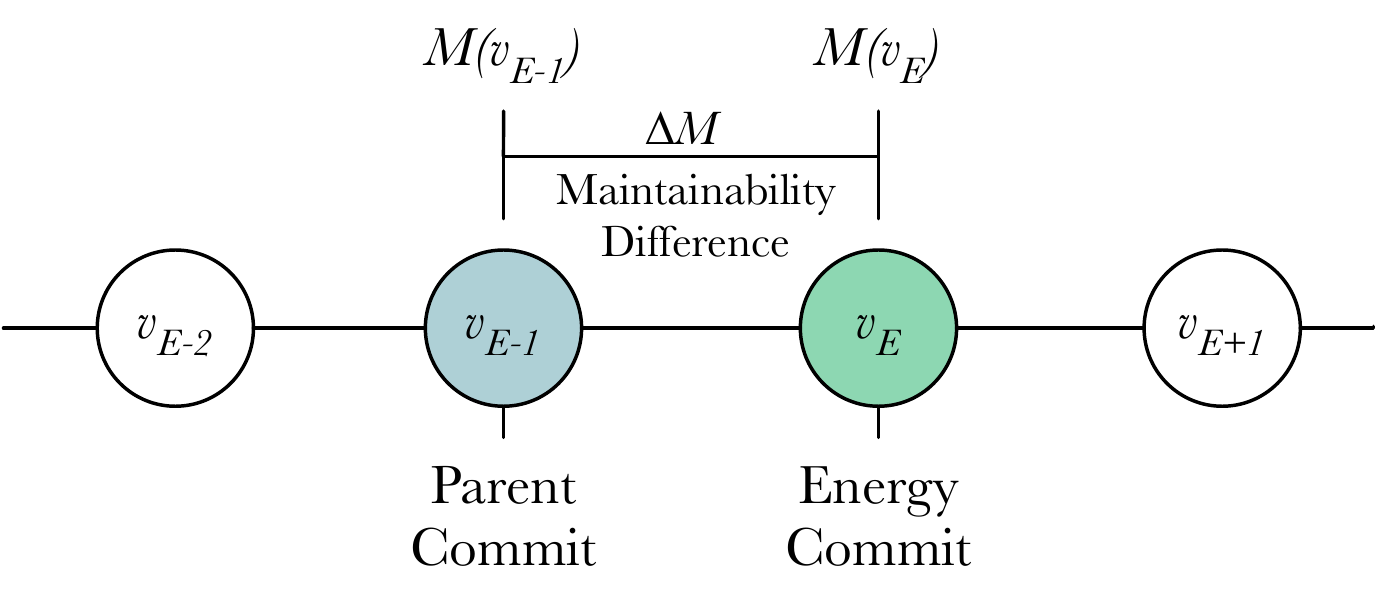}
  \caption{Maintainability difference for the energy commit $v_E$.}
  \label{fig:maintainability_diff}
\end{figure}

Since we want to analyze maintainability regression, we use BCH to
compute maintainability in two different versions of the Android app: a) the
version of the project before the energy commit ($v_{E-1}$) and b) the version
immediately after the energy commit ($v_E$). This is illustrated in
Fig.~\ref{fig:maintainability_diff}.

Although BCH provides a detailed report of the maintainability of the project,
it does not compute a final score that we can use to compare maintainability
amongst different projects. Thus, based in previous
work~\cite{olivari2018maintainable}, we designed an equation to capture the
distance between the current state of the project and the standard thresholds.
We have adjusted the equation to meet the following requirements:

\begin{itemize}

\item \textbf{The maintainability difference between two versions of the same
project is not affected by its size.} In this work, we want to evaluate the
identical energy patterns occurring in different projects. Thus, the metric
cannot use normalization based on its size -- we convert percentage data to
the respective number of lines of code.

\item \textbf{Distance to the thresholds in high severity levels is more
penalized than in low severity levels.} We use weights based on the severity
level to count lines of code that violate maintainability guidelines.
\end{itemize}

We compute the mean average of the maintainability score $M(v)$ for all the selected guidelines, as follows:
\begin{equation}
  M(v) = \sum_{g\in G}M_g(v)
\end{equation}
where:
\begin{conditions}
  G & selected maintainability guidelines from BCH (e.g., \emph{Write short units of code}, etc.)\\
  v & version of the app under analysis.\\
\end{conditions}

The maintenance $M$ based on the guideline $g$ for a given version of a project is computed with the following equation:
\begin{equation}
  M_g = \frac{1}{|L|}\sum_{l\in L}C(l),~ L=\{medium,high,veryHigh\}
\end{equation}
where:
\begin{conditions}
C   & compliance with the maintainability guideline for the given severity level (medium, high, and very high)\\
L   & severity levels of maintainability infractions.
\end{conditions}

The compliance $C$ for a given severity level $l$ is derived by:
\begin{equation}
  \label{eq:compliance}
  C(l) = LOC_{compliant}(l) - w(l)\cdot LOC_{\neg compliant}(l)
\end{equation}
where:
\begin{conditions}
LOC_{compliant}(l)      & lines of code that comply with the guideline at the given severity level $l$\\
LOC_{\neg compliant}(l) & lines of code that do not comply with the guideline at the given severity level $l$\\
w(l)                    & weight factor to boost the impact of non-compliant lines in comparison to compliant lines.\\
\end{conditions}

Finally, the term $w(l)$ is calculated as follows:

\begin{equation} w(l) = \frac{1 - T(l)}{T(l)} \end{equation} where:
\begin{conditions}
  T(l) & threshold in percentage of the lines of code that are
accepted to be non-compliant with the guideline for the severity level $l$.
This is a standard value defined by BCH, as illustrated in Fig.~\ref{fig:bch_noncompliant} and Table~\ref{tab:bch_report}.
\end{conditions}

In other words, the factor $w$ is used in Eq.~\ref{eq:compliance} to highlight
the lines of code that are not complying with the guideline. For instance, the
threshold for the severity level \emph{veryHigh} is defined in
Table~\ref{tab:bch_report} as $T(veryHigh)=6.9\%$, which derives to a weight of
$w(veryHigh)=13.5$. This means that, in this example, one non-compliant
guideline is decreasing maintainability score by $13.5$ points while a
compliant guideline is increasing by $1.0$ point. In addition, a version that
is perfectly aligned with the standard thresholds has a maintainability score
of zero.

Then, we compute the difference of maintainability ($\Delta M$) between the energy commit
($v_E$) and its parent commit ($v_{E-1}$), as illustrated in
Fig.~\ref{fig:maintainability_diff}.

\paragraph*{Statistical validation} To validate the maintainability differences
in different groups of commits (e.g., baseline and energy commits) we use the
Paired Wilcoxon signed-rank test with the significance level $\alpha=0.05$.
In other words, we test the null hypothesis that the maintainability difference
between pairs of versions $v_{E-1}$, $v_{E}$ (i.e., before and after an
energy-commit) follows a symmetric distribution around 0. This test does not
capture the absolute value of the maintainability differences. Thus, it is not
affected by confounding factors, such as the size of the code changes in
different groups.

To understand the effect-size, as advocated by the Common-language effect
sizes~\cite{mcgraw1992common}, we compute the mean difference, the median of
the difference, and the percentage of cases that reduce maintainability.

\subsection{Typical Maintainability Issues}

From the results collected in our dataset, we select the most evident examples
of maintainability issues that arise from improving energy efficiency. We
manually analyze these energy-oriented commits by examining its message and
code changes. The most evident cases are then discussed and presented to
illustrate common maintainability issues and bring awareness on how to avoid
common issues.

\section{Results}
\label{sec:results}

\begin{figure}[!t]
  \centering
  \includegraphics[width=\linewidth]{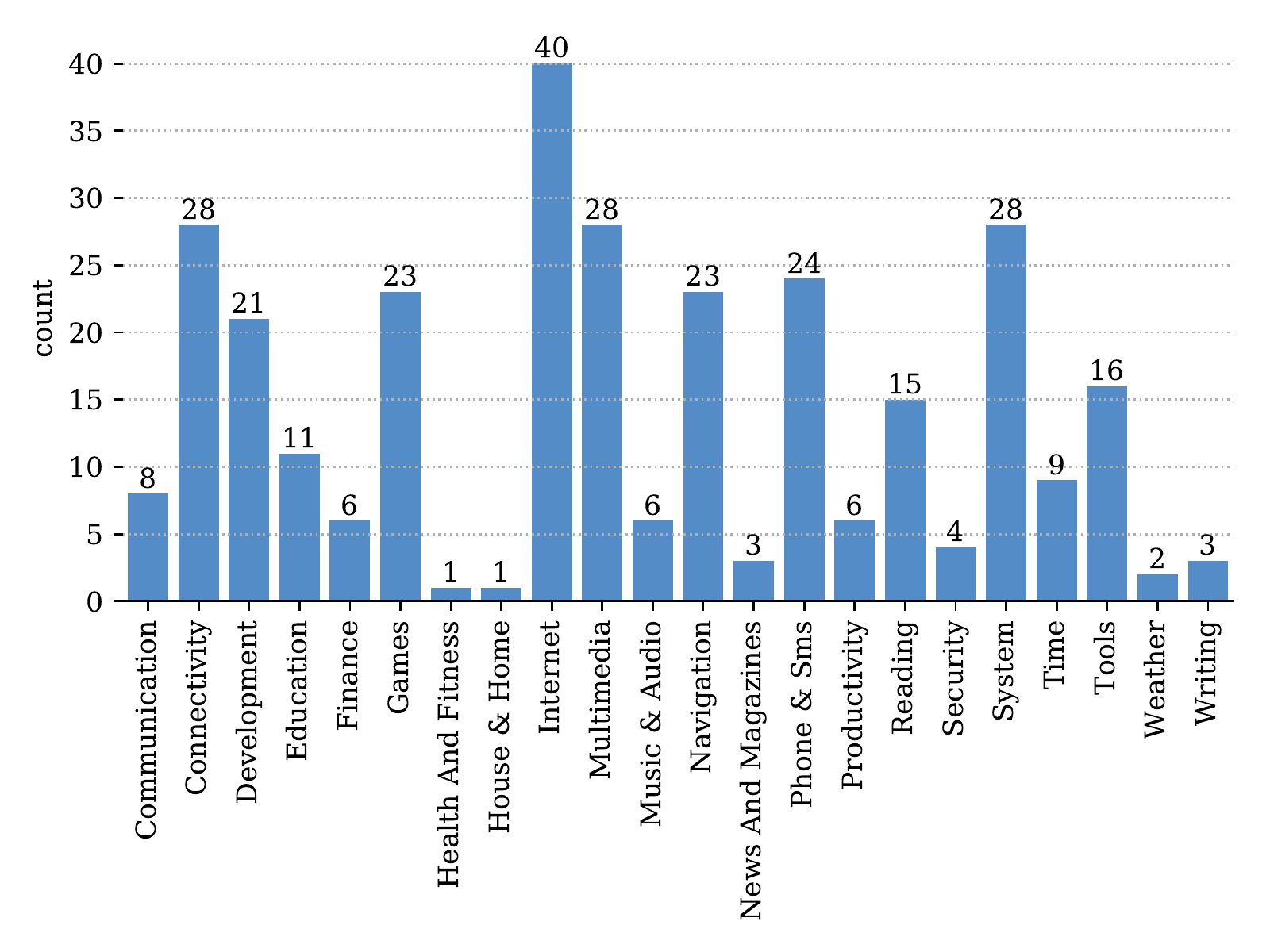}
  \caption{Categories of apps included in our study with the corresponding app count for each category.}
  \label{fig:app_categories}
\end{figure}

We evaluated a total of 539 energy commits and 539 baseline commits. These
commits comprise 306 apps distributed among 22 categories, as depicted in
Fig.~\ref{fig:app_categories}. In this section, we present the results for each proposed research question.

\rquestionSection{rq:maintainability}

\begin{figure}[!t]
  \centering
  \includegraphics[width=\linewidth]{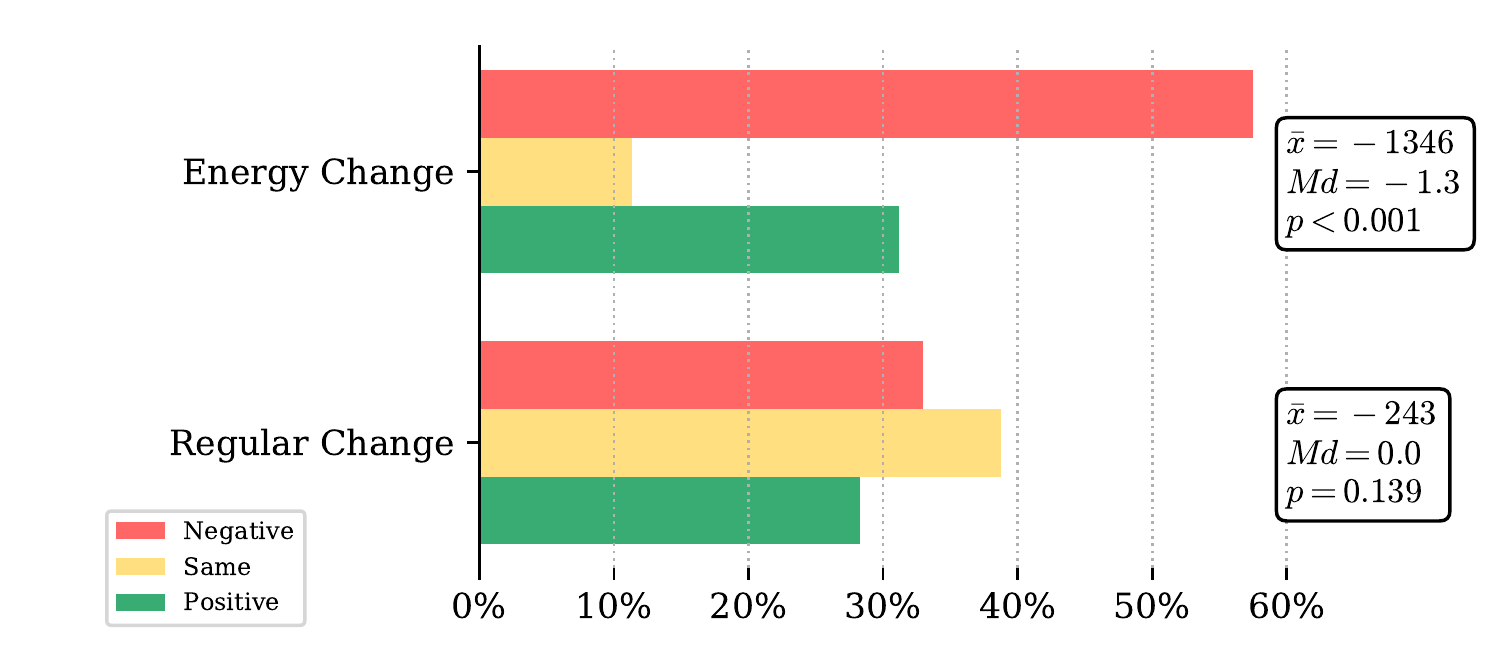}
  \caption{Maintainability differences for energy commits and baseline commits.}
  \label{fig:results_diffs}
\end{figure}

The results on the impact of different categories of commits in software
maintainability are presented in the plot bar of Fig.~\ref{fig:results_diffs}.
The plot presents the results for two groups of software changes:
\textbf{energy commits}, and \textbf{baseline commits}. For each group, the
figure provides three bars with the percentage of commits which 1) decrease
maintainability (on top, colored in red), 2) do not change maintainability (in
the middle, colored in yellow), and 3) increase maintainability (in the bottom,
colored in green). In addition, the figure provides, for each group, the mean
($\bar{x}$) and the median ($Md$) of the maintainability difference, and the
$p$-value of the Wilcoxon signed-rank test ($p$).

In the case of the regular commits, used as a baseline, $33.0\%$ decrease
maintainability ($183$ cases), $38.7\%$ do not change maintainability ($215$ cases),
and $28.3\%$ improve maintainability (157 cases). Since the p-value of the
Wilcoxon signed-rank test ($p = 0.139$) is not below the significance level
($\alpha = 0.05$), there is no statistical significance of the impact of
regular commits on maintainability.

On contrary, we observe clear changes for energy commits: $57.1\%$ ($310$ cases)
decrease software maintainability, $10.7\%$ do not change maintainability ($61$
cases), and $31.2\%$ improve maintainability ($168$ cases). The results for the Wilcoxon signed-rank test show statistical significance that energy
commits decrease the maintainability of Android applications ($p < 0.001$).

\rquestionSection{rq:patterns}

\begin{figure}[!t]
  \centering
  \includegraphics[width=\linewidth]{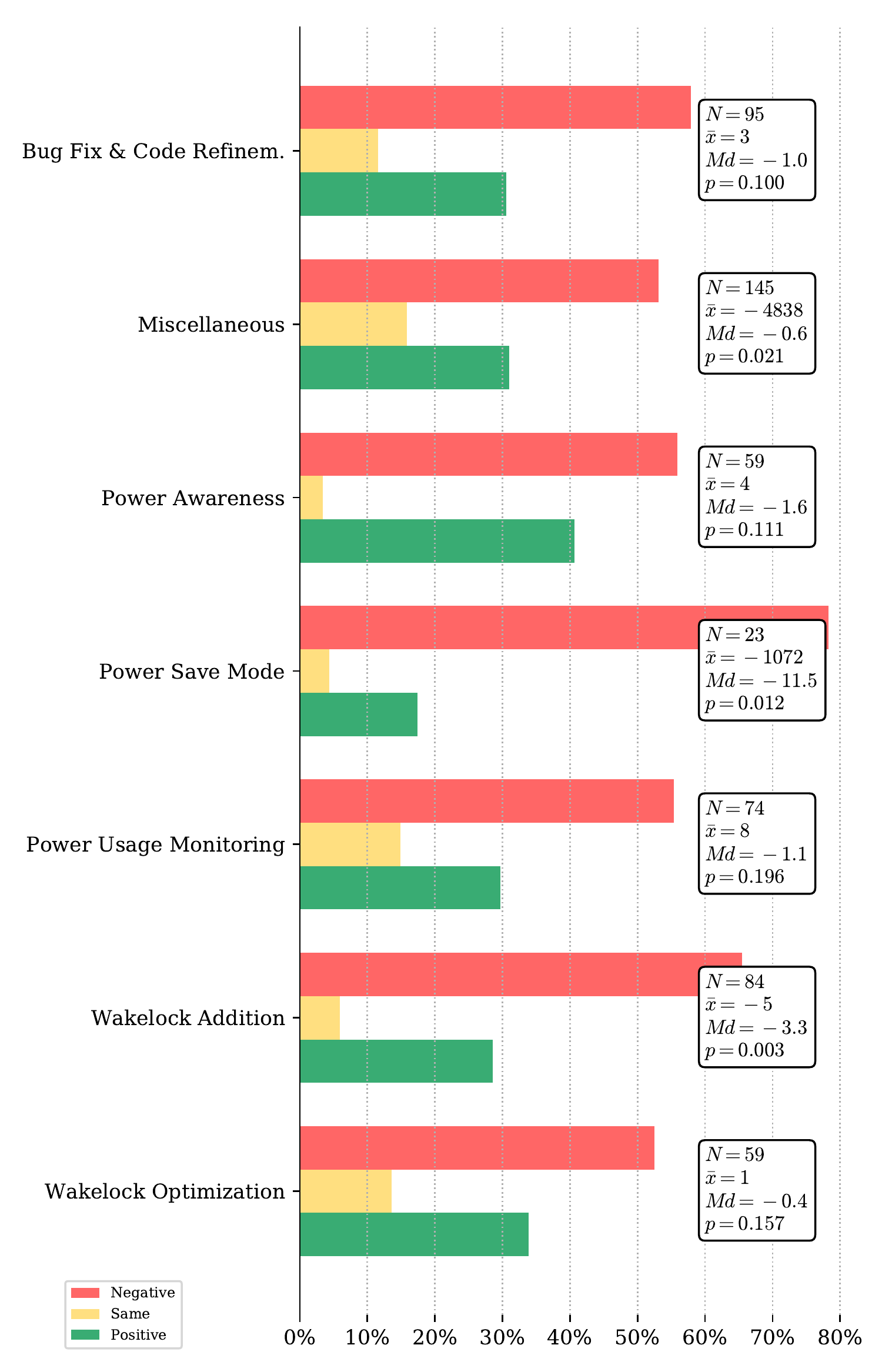}
  \caption{Maintainability differences among different types of energy commits.}
  \label{fig:patterns_diffs}
\end{figure}

Results of the maintainability impact per category of energy changes are
presented in Fig.~\ref{fig:patterns_diffs}. The Wilcoxon signed-rank test
yields statistical evidence that the categories \emph{Miscellaneous}
(\emph{p}~=~$0.021$), \emph{Power Save Mode} (\emph{p}~=~$0.012$), and \emph{Wakelock Addition}
(\emph{p}~=~$0.003$) significantly decrease the maintainability of Android projects.

The remaining patterns, (i.e., \emph{Bug Fix \& Code Refinement}, \emph{Power
Awareness}, \emph{Power Usage Monitoring}, and \emph{ Wakelock Optimization})
yielded more cases in which maintainability was negatively affected. However,
for these patterns, results are not statistically significant.

In the category \emph{Miscellaneous}, $53.1\%$ of changes ($77$ cases) have
decreased maintainability, while $15.9\%$ ($23$ cases) did not bring any impact,
and $31.0\%$ ($45$ cases) have improved maintainability. The impact is more
evident in the category \emph{Power Save Mode}, decreasing maintainability in
$78.3\%$ of changes ($18$ cases), leaving $4.3\%$ unaffected ($1$ case), and $17.4\%$
(4 cases) with an observed improvement in maintainability. Finally, in the
category \emph{Wakelock Addition}, $65.5\%$ have hindered maintainability ($55$
cases), $6.0\%$ ($5$ cases) have not yielded any difference, and $28.6\%$ ($24$
cases) have registered an improvement.

\rquestionSection{rq:examples}

The following examples illustrate a subset of the maintainability issues we
encounter that originate from energy-oriented changes (Maintainability
Instances \ref{bch:slide}--\ref{bch:last}).\footnote{The whole instances can be
found in the replication package:
\myurl{https://figshare.com/s/16397140e8183708d248}}

\bchInstance{ccrama/Slide}{070c2c6}{62f1a4b}{-949}%
{Merge two different categories of notifications in the same operation. This is
a common approach to improve energy efficiency, coined as \emph{Batch
Operations}\footnote{More information about \emph{Batch Operations} and other energy
pattern\myurl{https://tqrg.github.io/energy-patterns/\#/patterns/Batch\_Operation}.}~\cite{cruz2019catalog}.}%
{While coalescing different tasks, methods ended up being extremely large. As a
best practice, Java methods should not go over $15$ lines of
code~\cite{visser2016building}. Thus, the guideline \emph{Write Short Units of
Code} was violated in method \texttt{SubredditView.onOptionsItemSelected()},
which ended with $209$ lines of code. Several small helper methods should have
been implemented to keep this method short.}\label{bch:slide}

Maintainability Instance~\ref{bch:slide} shows an example of maintainability
issues that were likely introduced by the lack of awareness by developers on
best practices for maintainability. Before applying the code change, the
project already had $30$ methods with over $200$ lines of code. Extracting issues
that are strictly related to energy-efficiency improvements is not
straightforward. Thus, we skip examples in which this distinction was not
clear and opted for selecting maintainability issues that arise from improving
energy efficiency in projects with a positive maintainability score.

\bchInstance{mozilla/MozStumbler}{37819d9}{9d39be6}{-60}%
{New behavior to update the current GPS location. When the user is not
moving -- i.e., the accelerometer is not sensing any movement -- the GPS is
turned off and the location is assumed to be constant. When the user moves
again, the GPS location updater is reactivated. This instance is an example of
energy pattern and \emph{Sensor Fusion}~\cite{cruz2019catalog}.}%
{Although this new behavior for GPS updates was
added by default in the mobile application, the previous behavior remained as
an option in the codebase. This entailed some code duplications: the logic
needed to read data from GPS satellites is exactly the same in both behaviors.
This violates the \emph{Write Code Once} guideline.}\label{bch:first}\label{bch:fusion}

\bchInstance{mozilla/MozStumbler}{6ea0268}{6491bce}{-20}%
{Added support for a power save mode~\cite{cruz2019catalog},
in which the app stops scanning cell
towers and Wi-Fi networks. This change required adding extra logic in the \texttt{onCreate} method of \texttt{MainActivity} class. In short, the method was changed to verify whether the battery level was low and whether the \emph{Power Save Mode} was enabled in the app.}%
{Although the idea seems trivial, developers had to add 18 extra lines of code
to the already existing \texttt{MainActivity.onCreate()} method. The method
ended up with 45 lines of code, violating the \emph{Write Short Units of Code}
guideline.}

\bchInstance{einmalfel/PodListen}{2ed5a65}{35b93f7}{-20}%
{Add a preference in which users can opt to download new content (i.e.,
podcasts) only when the smartphone is connected to the charger. This is an
implementation of the energy patterns \emph{User Knows Best} and \emph{Power
Awareness}~\cite{cruz2019catalog}.%
}{%
By adding this new user-defined setting, conditional logic was added to the
beginning of the affected methods (e.g., method
\texttt{DownloadReceiver.updateDownloadQueue}) to verify the preferences and
the phone charging status. This leads to a higher number of branch points per
method (maximum recommended of four~\cite{visser2016building}), violating the
maintainability guideline \emph{Write Simple Units of Code}. In these cases,
the recommended approach is to split the method into simpler
ones.}\label{bch:powerAwareness}

\bchInstance{horn3t/PerformanceControl}{cb3080e}{34b881f}{-37}%
{Based on the battery level of the smartphone, adjust the power leveraged to
CPU and GPU cores. This a very low level code change that resorts to the
execution of bash commands to control the hardware of a smartphone device. This
example does not implement a documented energy pattern.}%
{Although the nature of the change implies adding code with poor readability,
there are other maintainability issues that should have been avoided. In
particular, the class \texttt{GPUClass}, which was added to control GPU power,
violates the guideline \emph{Separate Concerns in Modules}. The methods of this
class have a high number of references through the code (i.e., high fan-in). A
typical approach to address this issue is to split the class in separate
concerns\cite{visser2016building}.}\label{bch:last}

\section{Discussion}
\label{sec:discussion}

In this section, we answer our research questions, discussing the
implications of the analysis of results.

\rquestionSection{rq:maintainability}

\textbf{\emph{The majority of energy efficiency-oriented changes hinder the
maintainability of Android projects}}. Results presented in
Fig.~\ref{fig:results_diffs} shows a decrease in maintainability in 57\% of
the cases. This raises a new tradeoff when developers need to address
energy efficiency in their projects.

Previous work found evidence that developers struggle to improve the energy
efficiency of their software, lacking the knowledge and tools to aid in this
problem~\cite{pang2015programmers}. Our work corroborates by showing that
developers may have to reduce maintainability for the sake of energy efficiency.

In our perspective, developers need to be able to create energy efficient code
without potentially ruining the maintainability of their projects. Otherwise,
they may not apply such fixes or come with too many negative code maintenance
consequences. We understand that this problem needs to be addressed at several
levels:

\begin{itemize}

  \item \textbf{Mobile frameworks} need to feature energy patterns
  out-of-the-box without requiring too many changes in the software codebases.

  \item \textbf{Documentation} of mobile libraries and frameworks need to
  provide developers with the best practices to implement energy patterns.

  \item \textbf{Programming languages} should provide coding mechanisms to
  easily implement energy patterns without compromising maintainability.
  Previous work has already started addressing energy-efficiency concerns in
  programming languages~\cite{pereira2017energy,oliveira2017study}. Hopefully,
  these efforts can be ported to the official mobile programming languages
  (e.g., Java, Kotlin, Swift, etc.).

  \item \textbf{Mobile Developers} have to look out for maintainability issues
  when implementing energy patterns. Online services such as BCH, that play
  well in a continuous integration pipeline, can help developers to be more
  aware of the maintainability issues introduced by their changes. By bringing
  awareness, developers can put more effort on improving the maintainability
  of their code and avoid common issues (e.g., code duplication).

\end{itemize}

\rquestionSection{rq:patterns}

\emph{\textbf{Energy patterns \emph{Miscellaneous}, \emph{Power Save Mode}, and
\emph{Wakelock Addition} significantly decrease the maintainability of Android
projects.}} Although the remaining patterns \emph{Bug Fix \& Code Refinement},
\emph{Power Awareness}, \emph{Power Usage Monitoring}, and \emph{ Wakelock
Optimization} seem to reduce maintainability, no statistical evidence
was found.

This is particularly disconcerting because \emph{Power Save Mode} and
\emph{Wakelock Addition} are recommended as power management solutions in the
official documentation of the Android SDK\footnote{Documentation for
\emph{Power Save Mode} and \emph{Wakelocks}:
\myurl{https://developer.android.com/reference/android/os/PowerManager}.}.
However, it seems that more support is needed in order to implement patterns
without compromising the maintainability of Android projects.

Documentation should be enriched with more examples and best practices to
implement these patterns. We were not able to find those in the official
Android documentation. Moreover, the documentation does not consistently refer
to the \emph{Power Save Mode} pattern by this name, referring to it as
\emph{Battery Saver} in a few cases\footnote{Android
documentation using inconsistent names for \emph{Power Save Mode}:
\myurl{https://developer.android.com/about/versions/pie/power\#battery-saver}.}.

In addition, different Android versions feature different mechanisms to these
patterns. However, developers need to make sure their software runs efficiently
in different versions of Android~\cite{muccini2012software,an2018automatic}.
Thus, this requires adding specific logic for each API level, adding more
complexity to the code and making it less maintainable.

Along with the implications from RQ\ref{rq:maintainability}, we find that
improving support to \emph{Power Save Mode} and \emph{Wakelock Addition} would
immediately help developers ship maintainable and energy-efficient mobile
software. Actually, tools providing support to automatically apply these
patterns while preserving maintainability would be of great benefit.

\rquestionSection{rq:examples}

Although cases with the highest maintainability difference have clear
examples of bad maintainability, they are not entirely affected by the energy
improvement per se. That is, other factors, such as the low experience level of
the developer, may be the cause of the maintainability issues. This problem is
illustrated in the Maintainability Instance \ref{bch:slide}.

On the contrary, the examples presented in Maintainability Instances
\ref{bch:first}--\ref{bch:last} reveal maintainability issues that are
intrinsically related to the strategy used to improve energy efficiency. E.g.,
in the Maintainability Instance~\ref{bch:fusion}, developers created an additional
approach to collect sensor data but left the original one as an option. Since
the efficacy of the two approaches is different, developers decided to feature
both approaches in their app: one more effective but less efficient and the
other less effective but more efficient. Given that the app needs to run under
many different scenarios with different constraints, mobile apps often support
different approaches to the same feature. While this decision may be necessary,
the nature of these changes is prone to maintainability issues.

In the Maintainability Instance~\ref{bch:powerAwareness}, a number of
contextual pre-conditions related to the battery level of the smartphone were
checked before granting the execution of particular actions. Mobile development
frameworks should provide mechanisms to support typical battery-level scenarios
out of the box. For instance, using Java annotations,
particular actions could be postponed until power-related requirements are met.

Preliminary related work has proposed programming environments that address
energy-efficiency~\cite{yildirim2018ink,zhu2015programming}. We show that such
solutions are relevant in the context of mobile app development.
Moreover, related work has improved the specification of data types to select
the most energy-efficient type for a given
context~\cite{cohen2012energy,sampson2011enerj}. Nevertheless, these solutions
address energy efficiency decisions at low-level, lacking support for typical
design patterns to address the energy efficiency of mobile
apps\cite{cruz2019catalog}.

The analyzed examples show that maintainability issues lie mainly on the lack
of awareness by developers and the insufficient support of energy-efficiency
patterns from mobile platforms. New approaches ought to be delivered to help
developers assess the maintainability of their code changes when tackling
energy-efficiency requirements. For instance, continuous integration and
continuous development is a promising approach to address this issue. Although
it is known to promote software best practices~\cite{zhao2017impact}, they are
rare in the mobile app
world~\cite{cruz2019attention,kochhar2015understanding}. In addition, results
suggest that energy-related changes ought to be tackled by developers with
additional care (e.g., code reviews).

\section{Threats to Validity} \label{sec:t2v}

\vspace{-0.3em}
\subsection{Construct}
 
We use metrics derived from static code analysis to assess software
maintainability. However, this is a broad-scoped attribute that may not be
fully capture maintainability in its five sub-characteristics: modularity,
reusability, analyzability, modifiability, and testability. Nonetheless,
previous work has found high correlation between maintainability
sub-characteristics and BCH guidelines~\cite{bijlsma2012faster}.

In addition, different projects and contexts may require different
maintainability standards. Nonetheless, we use statistical hypothesis testing
to mitigate confounding factors. Moreover, BCH uses a representative
benchmark of closed and open source software systems to compute the thresholds
used in each maintainability
guideline\cite{visser2016building,baggen2012standardized}. This benchmark is
updated every year~\cite{visser2016building}.

\vspace{-0.3em}
\subsection{Internal}

Maintainability may be affected by different coding styles and
experience level from developers of the same project. We do not evaluate
differences at that level.
In addition, we do not evaluate the maintainability
difference for all regular commits in a project. Evaluating all the commits in
a project would not be feasible using our methodology. Thus, we assume that the
size of the dataset ($539$ commits) is enough to mitigate random variations in
the maintainability differences of the baseline.

The nature of baseline commits scopes general-purpose commits that may be different to
energy-oriented commits in a number of characteristics (e.g., lines of code).
We assure the two datasets are comparable by collecting the baseline set using a
random selection. Moreover, we do not analyze the maintainability difference
in terms of absolute values. In other words, we only evaluate whether the
maintainability was improved, not changed, or worsen. In future work, we plan
to address specific categories of changes in mobile apps.

\vspace{-0.3em}
\subsection{External}

The collection of energy-oriented commits used in this work comprises 
open source apps. Our methodology requires access to data that is not
publicly available for commercial apps. The extent to which this findings
generalize to commercial apps with non-open source licenses is not
assessed. Still, the maintainability challenges pinpointed in our work are
relevant to mobile app projects regardless of their license.

We only analyze Android apps. Different platforms and programming
languages may require different coding practices to address energy efficiency.
We did not study how our findings generalize to other mobile platforms.

We resort to a set of energy changes that were collected from four previous
works~\cite{moura2015mining,bao2016android,cruz2018measuring,cruz2019catalog}.
These works use the commit message provided by developers to classify a given
commit as an energy change. This approach discards energy changes that did not
have a commit message describing them as such. Since extending our
datasets to these commits is not trivial, we limit the scope of this study to
energy-oriented commits with an explicit commit message.
Finally, all the energy commits in this work are described in English.

\section{Related Work} \label{sec:rw}

In this section, we discuss related works on code maintainability, energy
efficiency patterns, and anti-pattern detection.

\vspace{-0.3em}
\subsection{Code maintainability}

Previous work has studied the evolution of maintainability issues during the
development of Android apps~\cite{malavolta2018maintainability}. The authors
have observed that maintainability decreases over time, being code duplication
the most common maintainability issue. In addition, they found evidence for the
fact that maintainability issues in Android apps occur independently of the
type of development activities performed by developers. Their work
uses a dataset from related work~\cite{pascarella2018self} with an
under-represented sample of energy activities, counting with only 12
occurrences. In this work, we focus on a larger sample, counting with 539
energy activities to analyze how energy activities affect the maintainability
of Android projects.

A use case study on the Java framework \emph{JHotDraw} suggests that the
adoption of design patterns is highly correlated with the maintainability of a
project -- i.e., the usage of design patterns do improve code
maintainability~\cite{hegedHus2012myth}. On contrary, related work shows that
some design patterns should be used with caution, since they may bring
maintainability and evolution issues to software
projects~\cite{khomh2008design}. Our work studies how these findings
apply in the case of energy patterns, for open-source Android apps.

Previous work studied the effect of programming languages in the quality of
code found~\cite{ray2014large}. It was found that language design has a
significant yet moderate impact on software quality. The authors have used the
number of defects as a construct of software quality. Our work analyzes
software quality in terms of code maintainability to study how it is affected
by energy efficiency-oriented commits.

\vspace{-0.3em}
\subsection{Energy efficiency patterns}

Previous works have studied the impact of different energy efficiency patterns
on mobile apps. Offloading heavy computation tasks to a cloud server was found
to reduce energy consumption up to 50\% in mobile
apps~\cite{kwon2013reducing}. Other patterns comprise featuring dark
user interface themes achieve better energy usage on mobile
devices~\cite{agolli2017investigating,linares2017gemma}. Other approaches have
improved energy efficiency by finding the optimal number of display updates in
a mobile app~\cite{kim2016content}. Another work has used regular
expression representations to assure an optimal usage of the energy intensive
resources of mobile devices~\cite{banerjee2016automated}.

The impact of logging practices of developers on the energy consumption of
Android apps has also been studied~\cite{chowdhury2018exploratory}.
From the 24 Android apps in this study, 19 exhibited at least one version in
which logging statements had a medium or large effect size on
energy consumption.

Energy patterns for mobile apps have been widely studied in the
literature~\cite{cruz2019catalog,bao2016android,moura2015mining}. Our work
acknowledges the importance of using energy patterns to improve energy
efficiency. However, we take a step further and study the impact of these
patterns on the quality of the app in terms of code maintainability. In
addition, we study the change-proneness of these techniques in mobile app
codebases.

\vspace{-0.3em}
\subsection{Detecting Anti-patterns in Mobile Apps}

Related work has studied how anti-patterns affect the overall energy
consumption of Android apps. Previous work on $60$ Android apps have studied the
influence of 9 Android-specific code smells on energy efficiency. Results
showed energy savings up to $87$ times after fixing all code
smells~\cite{palomba2019impact}. Another work has studied the impact of
eight performance-based code smells on the energy efficiency of mobile
apps~\cite{cruz2017performance}. It was found significant differences,
up to $5\%$, on energy consumption by fixing five of the studied code smells. Not
only code smells have been studied in this context. The impact of picture
smells on energy usage has also been assessed~\cite{carette2017investigating}.
It was found evidence that significant energy savings incur from using an
optimal image compression and format.

These works endorse the importance of using refactoring techniques to improve
energy efficiency. In fact, anti-pattern detectors and automatic refactoring
tools have been delivered to help developers ship energy efficient
code~\cite{morales2018earmo}. Cruz et al. have implemented an automatic
refactoring tool for Android apps to fix five performance issues that also
increase energy usage~\cite{cruz2018using}. Palomba et al. proposed an
automated tool to identify 15 Android-specific code
smells~\cite{palomba2017lightweight}. These code smells had been flagged by
previous work as a potential threat to the maintainability and the efficiency
of Android apps~\cite{reimann2014tool}. Our work differs by 1) identifying code
changes that hinder maintainability and 2) using code changes that already been
labeled has an energy improvement.

\section{Conclusion and Future Work}
\label{sec:conclusions}

In this work, we present an empirical study on the impact of energy commits on
the maintainability of mobile apps. We used the toolset BCH to collect
maintainability reports from a dataset of $539$ energy commits of open
source Android apps.

We have found evidence that energy-oriented commits significantly decrease
software maintainability in open source Android apps: $57\%$ of energy commits
were observed to reduce code maintainability. Conversely, no particular
influence on maintainability can be observed.
In particular, we show that the change on maintainability is more evident for
the patterns \emph{Power Save Mode} and \emph{Wakelock Addition}, in which
maintainability decreases in $78\%$ and $66\%$ of cases.

Our findings have direct implications for different stakeholders of mobile app
development. We highlight that mobile development frameworks
should provide mechanisms to implement energy patterns without hindering the
maintainability of mobile apps.

As future work, our empirical study can be extended in different ways: analyze
which maintainability guidelines are more affected from energy commits; analyze
how results stand for different categories of mobile apps; expand our
methodology with other software quality properties (e.g., reliability).
Furthermore, it would be interesting to validate our findings with other mobile
platforms (e.g., iOS), and also with desktop and server applications.

\section*{Acknowledgements}

\noindent We thank SIG's \emph{Better Code Hub} team for all the support as
well as help in validating our methodology.

This work is financed by National
Funds through the Portuguese funding agency, FCT - Fundação para a Ciência e a
Tecnologia with reference UID/EEA/50014/2019, the GreenLab
Project (ref. POCI-01-0145-FEDER-016718), and the FaultLocker Project (ref.
PTDC/CCI-COM/29300/2017).

\pagebreak

\balance
\bibliographystyle{IEEEtran}
\bibliography{bibliography}

\end{document}